\documentclass[slac_one]{revtex4}
\usepackage{graphicx}
\usepackage{fancyhdr}
\usepackage{axodraw}
\def\beq{\begin{equation}}
\def\eeq#1{\label{#1}\end{equation}}
\def\eeqn{\end{equation}}
\def\beqa{\begin{eqnarray}}
\def\eeqa#1{\label{#1}\end{eqnarray}}
\def\eeqan{\end{eqnarray}}
\def\CR{\nonumber \\ }
\def\leqn#1{(\ref{#1})}
\def\({\left(}
\def\){\right)}

\def\stacksymbols #1#2#3#4{\def\theguybelow{#2}
    \def\vp{\lower#3pt}
    \def\sp{\baselineskip0pt\lineskip#4pt}
    \mathrel{\mathpalette\intermediary#1}}

\def\intermediary#1#2{\vp\vbox{\sp
     \everycr={}\tabskip0pt
     \halign{$\mathsurround0pt#1\hfil##\hfil$\crcr#2\crcr
              \theguybelow\crcr}}}

\def\lapproxeq{\stacksymbols{<}{\sim}{2.5}{.2}}

\def\Mz{M_{\rm Z}}
\def\Mw{M_{\rm W}}

\newcommand{\bspace}{\!\!\!\!}
\newcommand{\met}{\mbox{$E{\bspace}/_{T}$}}

\def\mini#1{\leavevmode\hbox{\tiny #1}}

\def\g#1{g_{\mini{#1}}}

\def\to{\rightarrow}

\begin{document}

\title{{\small{2005 International Linear Collider Workshop - Stanford,
U.S.A.}}\\ 
\vspace{12pt}
Phenomenology of Higgsless Models at the LHC and the ILC}

\author{Andreas Birkedal\footnote{This talk was given by A.~Birkedal,
describing past and ongoing work performed in collaboration with the 
other authors.}, Konstantin T.~Matchev}
\affiliation{Institute for Fundamental Theory, University of Florida, 
Gainesville, FL 32611, USA}

\author{Maxim Perelstein}
\affiliation{Institute for High-Energy Phenomenology, Cornell University, 
Ithaca, NY 14853, USA}

\begin{abstract}
We investigate the signatures of the recently proposed Higgsless models 
at future colliders. We focus on tests of the mechanism of partial unitarity 
restoration in the longitudinal vector boson scattering, which do not depend 
on any Higgsless model-building details. We study the LHC discovery reach 
for charged massive vector boson resonances and show that
all of the preferred parameter space will be probed with $100\ {\rm fb}^{-1}$
of LHC data. We also discuss the prospects for experimental verification
of the Higgsless nature of the model at the LHC.  In addition, in this talk 
we present new results relevant for the discovery potential of Higgsless models 
at the International Linear Collider (ILC).
\end{abstract}

\maketitle

\thispagestyle{fancy}

\section{INTRODUCTION}
\label{sec:intro}

One of the greatest unsolved mysteries of the Terascale is the 
origin of electroweak symmetry breaking (EWSB).  
Within the usual description of the Standard Model (SM), a weakly
coupled Higgs boson performs this task.  However, it has still 
not been experimentally verified whether electroweak symmetry is 
broken by such a Higgs mechanism, by strong dynamics~\cite{TC}, 
or by something else.  This is one of the crucial questions particle 
physicists hope to answer in the upcoming experiments at the 
Large Hadron Collider (LHC) at CERN.

Experiments have already been able to put some constraints on 
theoretical ideas about EWSB.  
In theories involving EWSB by strong dynamics, the scale $\Lambda$ 
at which new physics enters can be guessed from the scale at which massive 
gauge boson scattering becomes non-unitary.  A simple estimate 
gives a value of 
\beq
\Lambda\sim4\pi M_W/g \sim 1.8~{\rm TeV},
\eeq{lambda}
which is disfavored 
by precision electroweak constraints (PEC)~\cite{PT}.  Thus, strong 
dynamics would seem to be largely ruled out as the source of EWSB.  
However, a new class of models, termed 
``Higgsless''~\cite{KK1,KK2,Nomura,KK3}, have been able to raise the 
scale of strong dynamics, allowing agreement with PEC~\cite{KKisOK}.

Realistic Higgsless models contain 
new TeV-scale weakly coupled states accessible at the LHC.
Among those, there are new massive vector bosons (MVB), 
heavy cousins of the $W$, $Z$ and $\gamma$ of the SM,
which are of primary interest. It is those states that delay 
unitarity violation and hence allow the scale $\Lambda$ to be 
raised~\cite{Chivukula:2001hz}.  Unfortunately, the details of 
the fermion sector of the theory are highly model-dependent.  
For instance, initial Higgsless models did not allow sufficient 
change in $\Lambda$ to agree with PEC~\cite{KKisnotOK,DHLR,PEWothers}, 
and modifications of the fermion sector were necessary.  However, 
the basic mechanism by which $\Lambda$ is raised is identical 
in all ``Higgsless'' models, even regardless of the number of 
underlying dimensions~\cite{decon}.  It is this mechanism that 
was studied in \cite{us}, focusing on its collider signatures. 
In this talk, we will review the analysis of Ref.~\cite{us}, and 
present some new results, relevant for the International Linear 
Collider (ILC).
In Sec.~\ref{sec:sumrules} we derive a set of sum rules which 
should be obeyed by the couplings between the new MVBs and the 
SM $W/Z$ gauge bosons. We identify discovery signatures of the 
new MVBs at the LHC which only rely on the couplings guaranteed 
by the sum rules, and compare to the SM Higgs search signals.
In Sec.~\ref{sec:lhc} we discuss the LHC reach for charged MVBs
and methods for testing the sum rules of Sec.~\ref{sec:sumrules}
in order to identify the ``Higgsless'' origin of the MVB resonances.
In Sec.~\ref{sec:ilc} we discuss the corresponding Higgsless
phenomenology at the ILC.

\section{UNITARITY SUM RULES}
\label{sec:sumrules}

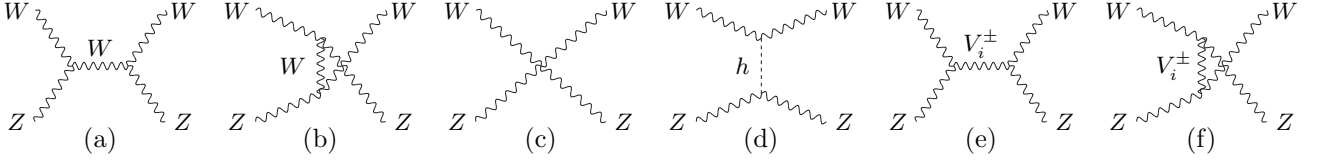
\begin{figure*}[t!]
\begin{center} 
{
\unitlength=0.7 pt
\SetScale{0.7}
\SetWidth{0.5}      
\normalsize    
{} \allowbreak
\begin{picture}(100,100)(0,0)
\Photon(15,80)(35,50){2}{6}
\Photon(15,20)(35,50){2}{6}
\Photon(35,50)(65,50){2}{6}
\Photon(65,50)(85,80){2}{6}
\Photon(65,50)(85,20){2}{6}
\Text(50,10)[c]{(a)}
\Text(50,60)[c]{\small $W$}
\Text( 5,80)[c]{\small $W$}
\Text( 5,20)[c]{\small $Z$}
\Text(95,80)[c]{\small $W$}
\Text(95,20)[c]{\small $Z$}
\end{picture}
\quad
\begin{picture}(100,100)(0,0)
\Photon(15,80)(50,65){2}{6}
\Photon(15,20)(50,35){2}{6}
\Photon(50,65)(50,35){2}{6}
\Photon(50,65)(85,20){2}{9}
\Photon(50,35)(85,80){2}{9}
\Text(50,10)[c]{(b)}
\Text(35,50)[c]{\small $W$}
\Text( 5,80)[c]{\small $W$}
\Text( 5,20)[c]{\small $Z$}
\Text(95,80)[c]{\small $W$}
\Text(95,20)[c]{\small $Z$}
\end{picture}\
\quad
\begin{picture}(100,100)(0,0)
\Photon(15,80)(85,20){2}{13}
\Photon(15,20)(85,80){2}{13}
\Text(50,10)[c]{(c)}
\Text( 5,80)[c]{\small $W$}
\Text( 5,20)[c]{\small $Z$}
\Text(95,80)[c]{\small $W$}
\Text(95,20)[c]{\small $Z$}
\end{picture}\
\quad
\begin{picture}(100,100)(0,0)
\Photon(15,80)(50,65){2}{6}
\Photon(15,20)(50,35){2}{6}
\DashLine(50,65)(50,35){2}
\Photon(50,65)(85,80){2}{6}
\Photon(50,35)(85,20){2}{6}
\Text(50,10)[c]{(d)}
\Text(40,50)[c]{\small $h$}
\Text( 5,80)[c]{\small $W$}
\Text( 5,20)[c]{\small $Z$}
\Text(95,80)[c]{\small $W$}
\Text(95,20)[c]{\small $Z$}
\end{picture}\
\quad
\begin{picture}(100,100)(0,0)
\Photon(15,80)(35,50){2}{6}
\Photon(15,20)(35,50){2}{6}
\Photon(35,50)(65,50){2}{6}
\Photon(65,50)(85,80){2}{6}
\Photon(65,50)(85,20){2}{6}
\Text(50,10)[c]{(e)}
\Text(50,62)[c]{\small $V_i^\pm$}
\Text( 5,80)[c]{\small $W$}
\Text( 5,20)[c]{\small $Z$}
\Text(95,80)[c]{\small $W$}
\Text(95,20)[c]{\small $Z$}
\end{picture}\
\quad
\begin{picture}(100,100)(0,0)
\Photon(15,80)(50,65){2}{6}
\Photon(15,20)(50,35){2}{6}
\Photon(50,65)(50,35){2}{6}
\Photon(50,65)(85,20){2}{9}
\Photon(50,35)(85,80){2}{9}
\Text(50,10)[c]{(f)}
\Text(35,50)[c]{\small $V_i^\pm$}
\Text( 5,80)[c]{\small $W$}
\Text( 5,20)[c]{\small $Z$}
\Text(95,80)[c]{\small $W$}
\Text(95,20)[c]{\small $Z$}
\end{picture}
}
\end{center}
\caption{Diagrams contributing to the $W^\pm Z\to W^\pm Z$ scattering process: 
(a), (b) and (c) appear both in the SM and in
Higgsless models, (d) only appears in the SM,
while (e) and (f) only appear in Higgsless models.}
\label{wzwz}
\end{figure*}

Consider the elastic scattering process $W^\pm_L Z_L\to W^\pm_L Z_L$. 
In the absence of the Higgs boson, this process receives contributions 
from the three Feynman diagrams shown in Figs.~\ref{wzwz}(a)--(c). 
The resulting amplitude contains terms which grow with the energy $E$
of the incoming particle as
$E^4$ and $E^2$ and ultimately cause violation of unitarity at sufficiently
high energies. In the SM, both of these terms are precisely cancelled by the
contribution of the Higgs exchange diagram in Fig.~\ref{wzwz}(d).
In the Higgsless theories, on the other hand, the diagram of Fig.~\ref{wzwz}(d)
is absent, and the process instead receives additional contributions from 
the diagrams in Figs.~\ref{wzwz}(e) and \ref{wzwz}(f), where $V_i^\pm$ denotes the 
charged MVB of mass $M_i^\pm$. The index $i$ corresponds to the KK 
level of the state in the case of a 5D theory, or labels the 
mass eigenstates in the case of a 4D deconstructed theory. 
Remarkably, the $E^4$ and $E^2$ terms can again be exactly cancelled 
by the contribution of the MVBs, provided that the following 
sum rules are satisfied~\cite{us}:
\beqa
\g{WWZZ} &=& \g{WWZ}^2 \,+\, \sum_i (\g{WZV}^{(i)})^2, \CR
2(\g{WWZZ}-\g{WWZ}^2)(\Mw^2+\Mz^2) + \g{WWZ}^2\,\frac{\Mz^4}{\Mw^2} 
&=& \sum_i (\g{WZV}^{(i)})^2\,
\left[3(M^\pm_i)^2-\frac{(\Mz^2-\Mw^2)^2}{(M^\pm_i)^2}\right]. 
\eeqa{sumW} 
Here $M_W$ ($M_Z$) is the $W$-boson ($Z$-boson) mass and the notation for 
the triple and quartic gauge boson couplings is self-explanatory.
In 5D theories, these equations are satisfied exactly if all the KK states, 
$i=1\ldots\infty$, are taken into account. This is not an accident, but a 
consequence of the gauge symmetry and locality of the underlying theory. 
While this is not sufficient to ensure unitarity at all energies 
(the increasing number of inelastic channels ultimately results 
in unitarity violation), the strong coupling scale can be 
significantly higher than the naive estimate~\leqn{lambda}. 
For example, in the warped-space Higgsless models~\cite{KK2,KKisOK} 
unitarity is violated at the scale~\cite{Pap} 
\beq
\Lambda_{\rm NDA} \sim \frac{3\pi^4}{g^2}\frac{\Mw^2}{M^\pm_1},
\eeq{NDA}
which is typically of order 5--10 TeV. In 4D models, the number of 
MVBs is finite and the second of the sum rules~\leqn{sumW} is 
only satisfied approximately; however, our numerical study of sample models 
indicates that the violation of the sum rule has to be very small 
(at the level of 1\%) to achieve an adequate improvement in $\Lambda$. 

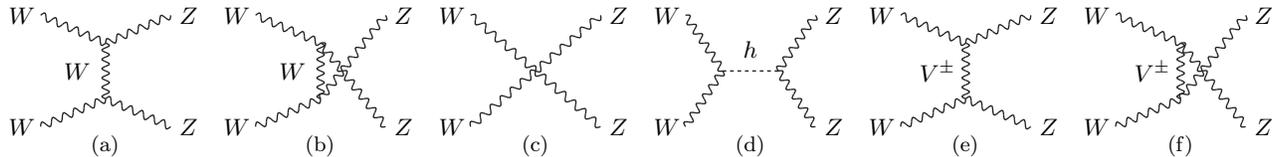
\begin{figure*}[t]
\begin{center} 
{
\unitlength=0.7 pt
\SetScale{0.7}
\SetWidth{0.7}      
\footnotesize    
{} 
\allowbreak
%
\begin{picture}(100,100)(0,0)
\Photon(15,80)(50,65){2}{6}
\Photon(15,20)(50,35){2}{6}
\Photon(50,65)(50,35){2}{6}
\Photon(50,65)(85,80){2}{6}
\Photon(50,35)(85,20){2}{6}
\Text(50,10)[c]{(a)}
\Text(35,50)[c]{\small $W$}
\Text( 5,80)[c]{\small $W$}
\Text( 5,20)[c]{\small $W$}
\Text(95,80)[c]{\small $Z$}
\Text(95,20)[c]{\small $Z$}
\end{picture}\
\quad
\begin{picture}(100,100)(0,0)
\Photon(15,80)(50,65){2}{6}
\Photon(15,20)(50,35){2}{6}
\Photon(50,65)(50,35){2}{6}
\Photon(50,65)(85,20){2}{9}
\Photon(50,35)(85,80){2}{9}
\Text(50,10)[c]{(b)}
\Text(35,50)[c]{\small $W$}
\Text( 5,80)[c]{\small $W$}
\Text( 5,20)[c]{\small $W$}
\Text(95,80)[c]{\small $Z$}
\Text(95,20)[c]{\small $Z$}
\end{picture}\
\quad
\begin{picture}(100,100)(0,0)
\Photon(15,80)(85,20){2}{13}
\Photon(15,20)(85,80){2}{13}
\Text(50,10)[c]{(c)}
\Text( 5,80)[c]{\small $W$}
\Text( 5,20)[c]{\small $W$}
\Text(95,80)[c]{\small $Z$}
\Text(95,20)[c]{\small $Z$}
\end{picture}\
\quad
%
\begin{picture}(100,100)(0,0)
\Photon(15,80)(35,50){2}{6}
\Photon(15,20)(35,50){2}{6}
\DashLine(35,50)(65,50){2}
\Photon(65,50)(85,80){2}{6}
\Photon(65,50)(85,20){2}{6}
\Text(50,10)[c]{(d)}
\Text(50,62)[c]{\small $h$}
\Text( 5,80)[c]{\small $W$}
\Text( 5,20)[c]{\small $W$}
\Text(95,80)[c]{\small $Z$}
\Text(95,20)[c]{\small $Z$}
\end{picture}\
\quad
%
\begin{picture}(100,100)(0,0)
\Photon(15,80)(50,65){2}{6}
\Photon(15,20)(50,35){2}{6}
\Photon(50,65)(50,35){2}{6}
\Photon(50,65)(85,80){2}{6}
\Photon(50,35)(85,20){2}{6}
\Text(50,10)[c]{(e)}
\Text(35,50)[c]{\small $V^\pm$}
\Text( 5,80)[c]{\small $W$}
\Text( 5,20)[c]{\small $W$}
\Text(95,80)[c]{\small $Z$}
\Text(95,20)[c]{\small $Z$}
\end{picture}\
\quad
%
\begin{picture}(100,100)(0,0)
\Photon(15,80)(50,65){2}{6}
\Photon(15,20)(50,35){2}{6}
\Photon(50,65)(50,35){2}{6}
\Photon(50,35)(85,80){2}{9}
\Photon(50,65)(85,20){2}{9}
\Text(50,10)[c]{(f)}
\Text(35,50)[c]{\small $V^\pm$}
\Text( 5,80)[c]{\small $W$}
\Text( 5,20)[c]{\small $W$}
\Text(95,80)[c]{\small $Z$}
\Text(95,20)[c]{\small $Z$}
\end{picture}\
}
\end{center}
\caption{Diagrams contributing to the $W^\pm W^\mp\to ZZ$ scattering process: 
(a), (b)  and (c) appear both in the SM and in Higgsless models, 
(d) only appears in the SM, and 
(e) and (f) only appear in Higgsless models.
}
\label{wwzz}
\end{figure*}

Considering the $W^+_LW^-_L\to  W^+_LW^-_L$ scattering process yields sum 
rules constraining the couplings of the neutral MVBs $V_i^0$
(with masses denoted by $M^0_i$) \cite{KK1}:
\beqa
\g{WWWW} &=& \g{WWZ}^2 + \g{WW$\gamma$}^2 \,+\, \sum_i (\g{WWV}^{(i)})^2, \CR
4\g{WWWW}\,\Mw^2 &=& 3\,\left[\g{WWZ}^2 \Mz^2 + \sum_i (\g{WWV}^{(i)})^2
\,(M^0_i)^2\right]\, .
\eeqa{sumZ} 
Considering other channels such as $W_L^+W_L^-\to ZZ$ (see Fig.~\ref{wwzz})
and $ZZ\to ZZ$ does not yield any new sum rules.
The presence of multiple MVBs, whose couplings obey Eqs.~\leqn{sumW},~\leqn{sumZ}, 
is a generic prediction of the Higgsless models.

Our study of the collider phenomenology 
in the Higgsless models will focus on the vector boson fusion processes. 
These processes are attractive for two reasons. Firstly, the production 
of MVBs via vector boson fusion is relatively model-independent, since 
the couplings are constrained by the sum rules~\leqn{sumW},~\leqn{sumZ}. 
This is in sharp contrast with the 
Drell-Yan production mechanism~\cite{DHLR}, which dominates for the 
conventional $W^\prime$ and $Z^\prime$ bosons but is likely to be 
suppressed for the Higgsless MVBs due to their small couplings to 
fermions, as needed to evade PEC~\cite{KKisOK}. In the following,
unless specified otherwise, 
we shall assume that the MVBs have no appreciable couplings to SM fermions.
Secondly, if enough couplings and masses can be measured, 
these processes can provide a {\em test} of the sum rules, probing 
the mechanism of partial unitarity restoration.

Eq.~\leqn{NDA} indicates that the first MVB should appear 
below $\sim 1$ TeV, and thus be accessible at the LHC. 
For $V^\pm_1$, the sum rules~\leqn{sumW} imply an inequality
\beq
\g{WZV}^{(1)}  
\lapproxeq \frac{\g{WWZ}\Mz^2}{\sqrt{3}M^\pm_1\Mw}.
\eeq{bound}
This bound is quite stringent ($\g{WZV}^{(1)}\lapproxeq0.04$ 
for $M^\pm_1=700$ GeV). 
Also, convergence of the sum rules~\leqn{sumW} requires 
$
\g{WZV}^{(k)}\,\propto\,k^{-1/2}\,(M^\pm_k)^{-1}.
$
The combination of heavier masses and lower couplings means that 
the heavier MVBs may well be unobservable, so that only the 
$V_1$ states can be studied. The "saturation limit", in which there 
is only a single set of MVBs whose couplings saturate the sum 
rules, is likely to provide a good approximation to the 
phenomenology of the realistic Higgsless models. 
In this limit, the partial width of the $V^\pm_1$ is given by
\begin{equation}
\Gamma(V^\pm_1\to W^\pm Z)\approx \frac{\alpha\ (M^\pm_1)^3}{144\, \sin^2\theta_W\, M_W^2}\, .
\label{width}
\end{equation}

\begin{figure*}[t]
\centering
\includegraphics[width=85mm]{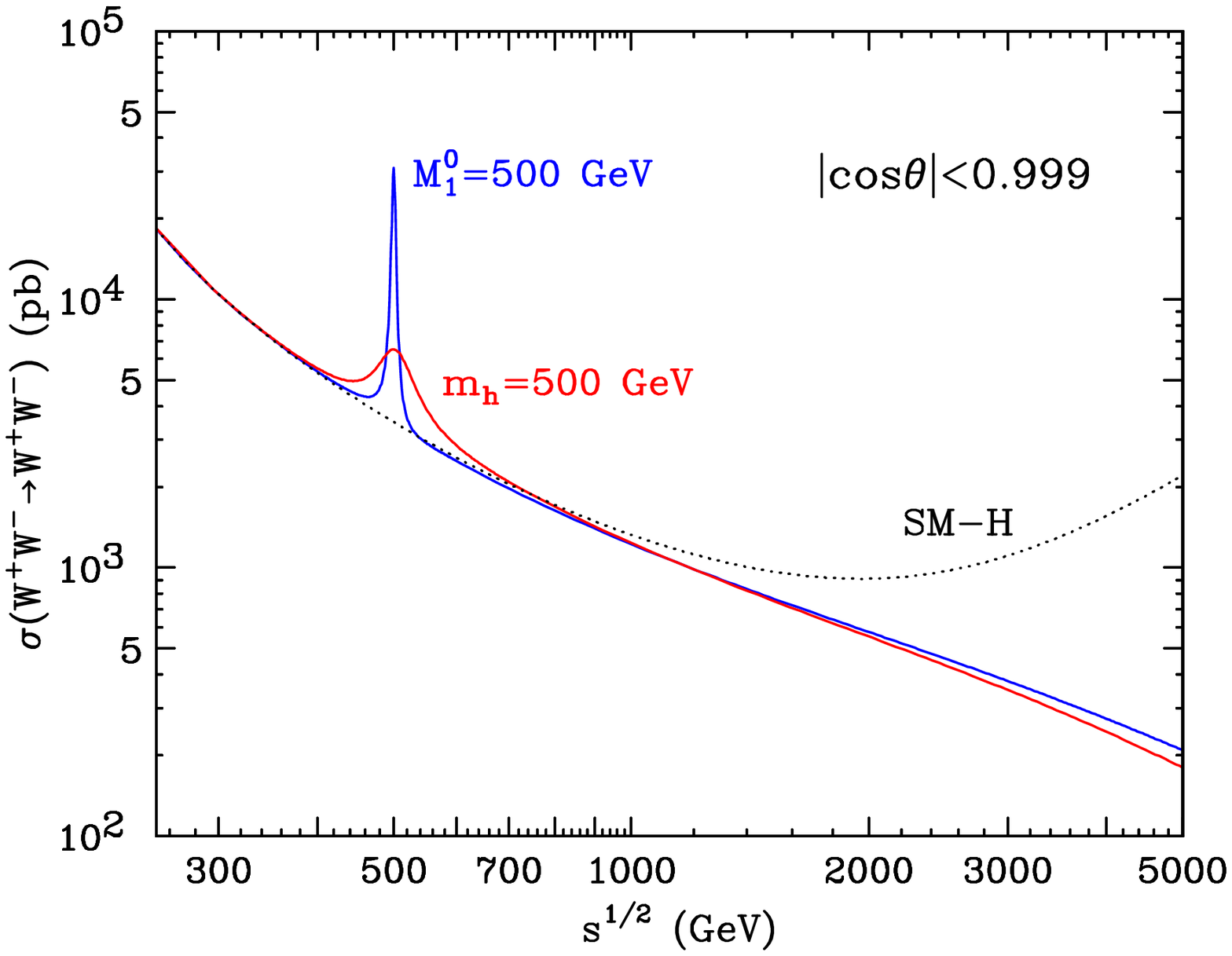}
\includegraphics[width=85mm]{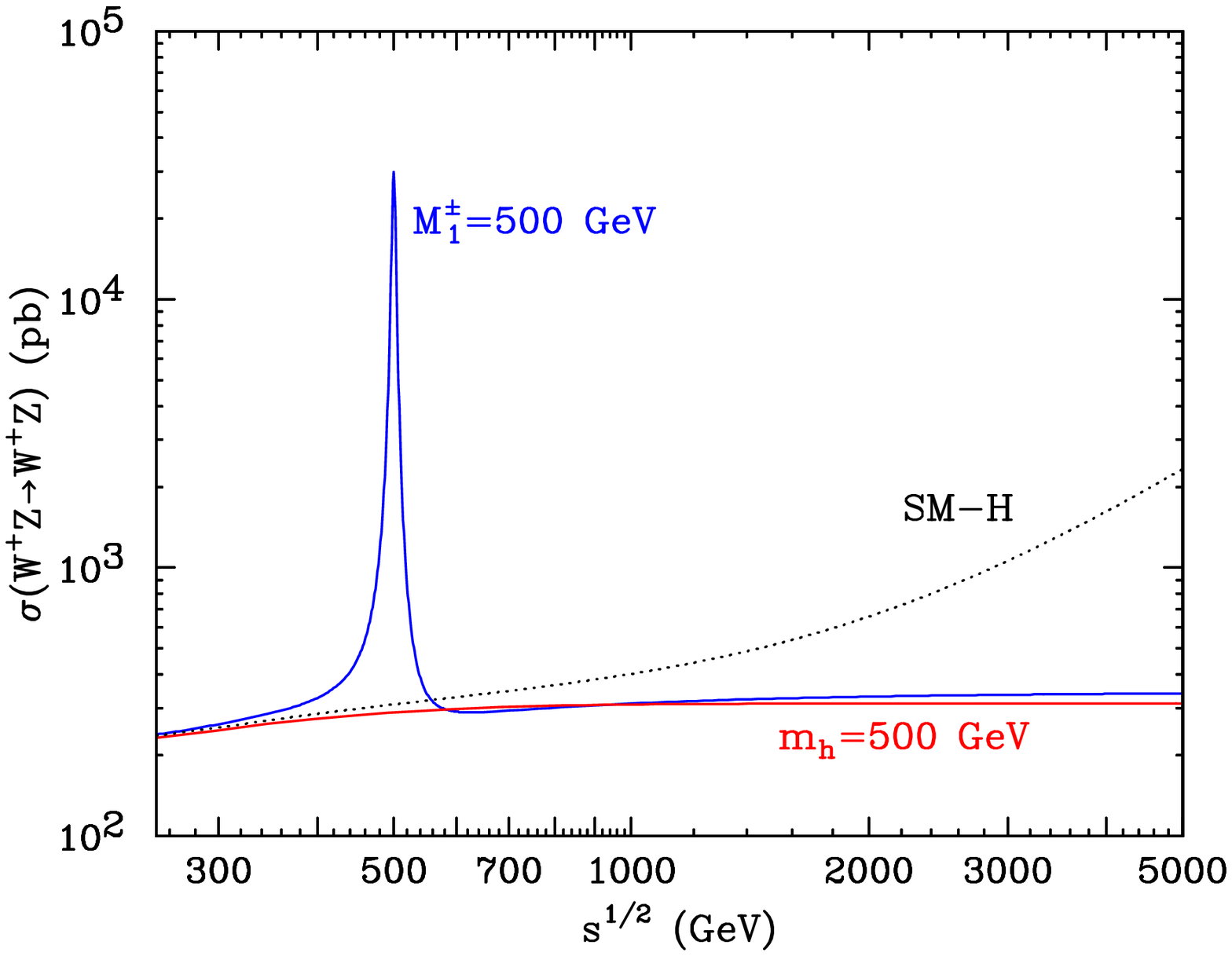}
\caption{Elastic scattering cross-sections for $WW\to WW$ (left) and
$WZ\to WZ$ (right) in the SM without a Higgs boson (SM-H) (dotted), 
the SM with a 500 GeV Higgs boson (red) and
the Higgsless model with a 500 GeV MVB (blue).} 
\label{fig:parton}
\end{figure*}

Given the couplings of the MVBs to the SM $W$ and $Z$, we can now 
predict (at the parton level) the size of the new physics signals
in the various channels of vector boson fusion. Fig.~\ref{fig:parton}
provides an illustration for the case of $WW\to WW$ and $WZ\to WZ$. 
We show the expected signal for either a SM Higgs boson of mass $m_h=500$ GeV, 
or the corresponding MVB $V_1$ of mass $500$ GeV in the saturation limit.
The sum rules (\ref{sumZ}) govern the signal in the $WW\to WW$ channel
shown in the left panel of Fig.~\ref{fig:parton}. However, the $WW$ final
state is difficult to observe over the SM backgrounds at the LHC: 
in the dilepton channel there is no resonance structure, while 
the jetty channels suffer from large QCD backgrounds. It is therefore 
rather challenging to test the sum rules (\ref{sumZ}). Notice that 
even if a $WW$ resonance is observed, without a test of the sum rules
(\ref{sumZ}), its interpretation is unclear, since the SM Higgs boson
is {\em also} expected to appear as a $WW$ resonance 
(see the left panel in Fig.~\ref{fig:parton}).

We shall therefore concentrate on the $WZ\to WZ$ channel, in which
the Higgsless model predicts a series of resonances 
as in Fig.~\ref{wzwz}(e), while in the SM the amplitude is 
unitarized by the $t$-channel diagram of Fig.~\ref{wzwz}(d) 
and has no resonance (see the right panel in Fig.~\ref{fig:parton}). 
Conventional theories of 
EWSB by strong dynamics may also contain a resonance in this channel, 
but it is likely to be heavy ($\sim 2$ TeV for QCD-like theories) 
and broad due to strong coupling. In contrast, the MVB resonance
is very narrow, as can be seen from Fig.~\ref{fig:parton} and 
Eq.~(\ref{width}). In fact it is almost a factor of 20 narrower 
than a SM Higgs boson of the same mass. This is primarily due to the 
vector nature of the MVB and our assumption that it only has 
a single decay channel. 
We therefore conclude that a resonance in the $WZ\to WZ$ channel 
would be a smoking gun for the Higgsless model (for alternative 
interpretations involving extended Higgs sectors, 
see \cite{Asakawa:2005gv} and references therein). 
Finally, the $WW\to ZZ$ channel is a good discriminator as well, 
since it will exhibit a resonance for the case of the 
SM but not the Higgsless models (see Fig.~\ref{wwzz}). 
A comparison of the resonant structure of the three vector boson 
fusion final states is shown in Table~\ref{tab:res}.

\begin{table}[htb]
\begin{center}
\caption{Comparison of the resonance structure of the SM and Higgsless models 
in different vector boson fusion channels.}
\vspace{5mm}
\begin{tabular}{||c||c|c|c||}
\hline\hline
Model
& $WW\to WW$
& $WZ\to WZ$
& $WW\to ZZ$ \\ 
\hline\hline
SM
& Yes
& No
& Yes \\ 
\hline
Higgsless
& Yes
& Yes
& No  \\ 
\hline\hline
\end{tabular}
\label{tab:res}
\end{center}
\end{table}

\section{COLLIDER PHENOMENOLOGY AT THE LHC}
\label{sec:lhc}

\begin{figure*}[t]
\centering
\includegraphics[width=85mm]{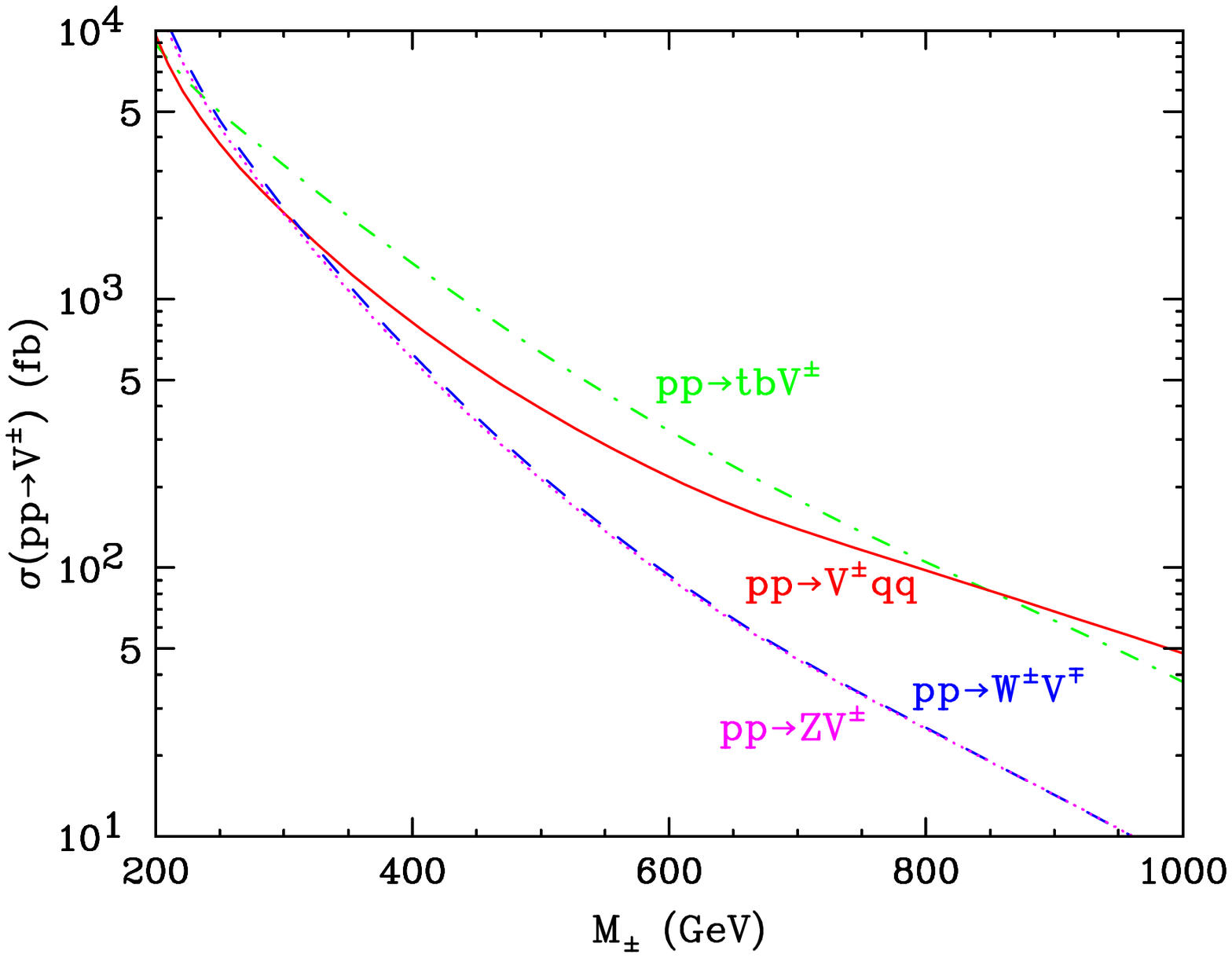}
\includegraphics[width=85mm]{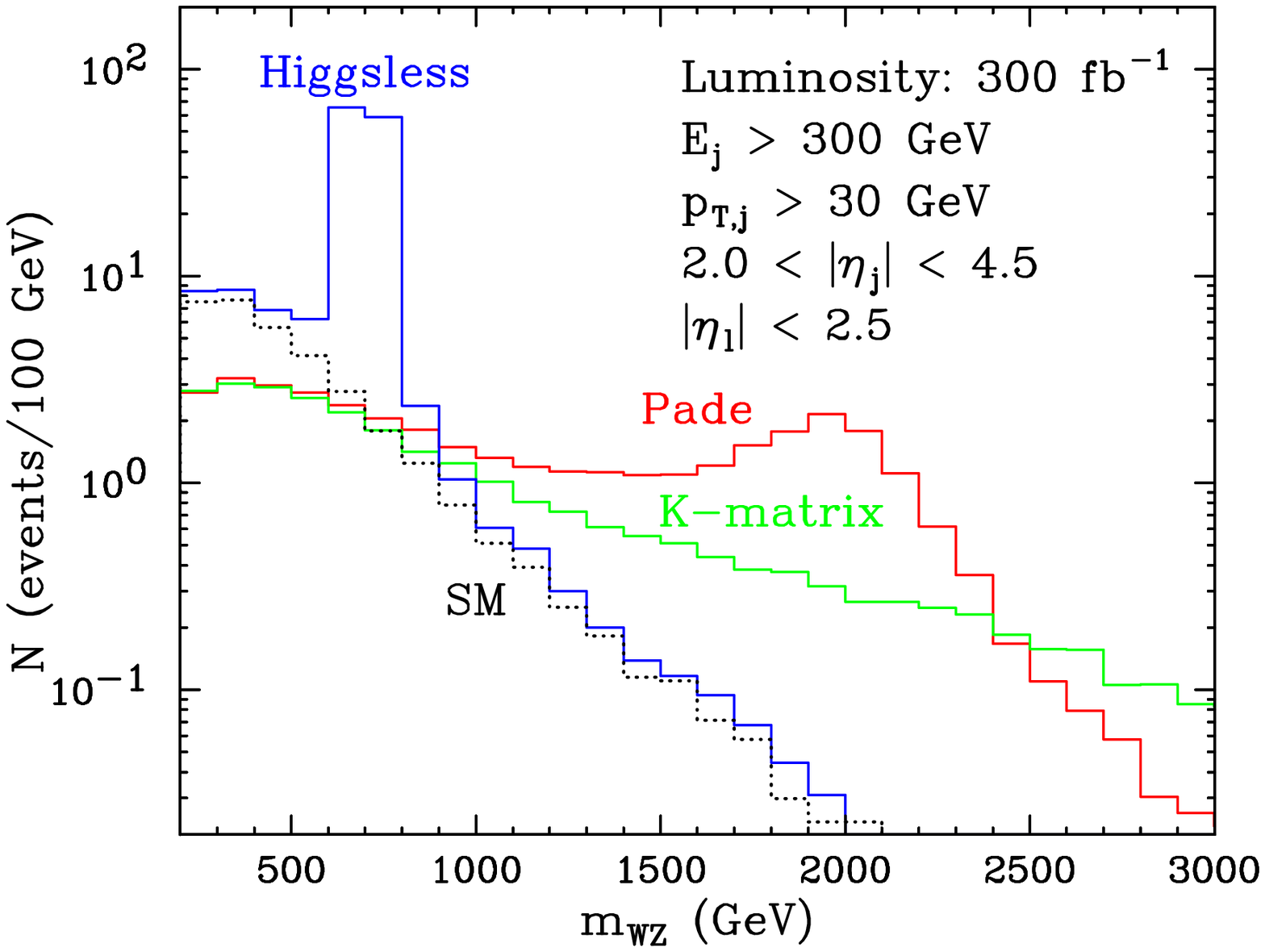}
\caption{Left: Production cross-sections of $V^\pm$ at the LHC. Here 
$tbV^\pm$ production assumes SM-like couplings to third generation quarks.
Right: The number of events per 100 GeV bin in the $2j+ 3\ell+\nu$ 
channel at the LHC with an integrated luminosity of 300
fb$^{-1}$ and cuts as indicated in the figure. Results are shown for
the SM (dotted), the Higgsless model with $M^\pm_1=700$ GeV (blue), 
and two "unitarization" models: Pad\'e (red) and K-matrix (green)
\cite{DH}.} 
\label{fig:lhc}
\end{figure*}

At the LHC, the vector boson fusion processes will occur as a result 
of $W/Z$ bremsstrahlung off quarks. The typical final state for such 
events includes two forward jets in addition to a pair of gauge bosons. 
The production cross section of $V^\pm_1$ in association with two jets
is shown by the solid line in the left panel of Fig.~\ref{fig:lhc}.
To estimate the prospects for the charged MVB search at the LHC, we 
require that both jets be observable (we assume jet rapidity coverage 
of $|\eta|\leq 4.5$), and impose the following lower cuts on 
the jet rapidity, energy, and transverse momentum: 
$|\eta|>2, E>300$ GeV, $p_T>30$ GeV. These requirements 
enhance the contribution of the vector boson fusion
diagrams relative to the irreducible background of the non-fusion 
$q\bar{q}'\to WZ$ SM process as well as $q\bar{q}'\to V_1^\pm$ Drell-Yan
process. The ``gold-plated'' final state~\cite{Bagger:1993zf}
for this search is $2j+3\ell$+\met, with the 
additional kinematic requirement that two of the leptons have to be consistent 
with a $Z$ decay. We assume lepton rapidity coverage of
$|\eta|<2.5$. The $WZ$ invariant mass, $m_{WZ}$, 
can be reconstructed using the missing transverse energy measurement and 
requiring that the neutrino and the odd lepton form a $W$. 
The number of "gold-plated" events (including all lepton sign combinations)
in a 300 fb$^{-1}$ LHC data sample, 
as a function of $m_{WZ}$, is shown in Fig.~\ref{fig:lhc},
for the SM (dotted), the Higgsless model with $M^\pm_1=700$ GeV (blue), 
and two "unitarization" models: Pad\'e (red) and K-matrix (green)~\cite{DH}
(for details, see \cite{us}).  
The Higgsless model can be easily identified by observing the 
MVB resonance: for the chosen parameters, the dataset contains 
$130$ $V_1^\pm\to W^\pm Z \to 3\ell+\nu$ events. The irreducible 
non-fusion SM background is effectively suppressed by the cuts: 
the entire dataset shown in Fig.~\ref{fig:lhc} contains only 
$6$ such events. We therefore estimate the discovery reach 
for $V^\pm_1$ resonance by requiring 10 signal events after cuts.
The efficiency of the cuts for $500\ {\rm GeV}\le M^\pm_1\le 3\ {\rm TeV}$
is in the range $20-25\%$.
We then find that with $10\ {\rm fb}^{-1}$ of data, corresponding 
to 1 year of running at low luminosity, the LHC will probe the Higgsless models 
up to $M^{\pm}_1\lapproxeq 550$ GeV, while covering the whole 
preferred range up to $M^{\pm}_1=1$ TeV requires $60\ {\rm fb}^{-1}$.
Note, however, that one should expect a certain amount of 
reducible background with fake and/or non-isolated leptons.

Once the $V^\pm_1$ resonance is discovered, identifying it as part of 
a Higgsless model requires testing the sum rules (\ref{sumW}) by
measuring its mass $M^\pm_1$ and coupling $g^{(1)}_{WZV}$. 
The coupling can be determined from the total $V^\pm_1$
production cross section $\sigma_{\rm tot}$.
However, we are observing the $V^\pm_1$ resonance in an exclusive channel,
which only yields the product $\sigma_{\rm tot}\,BR(V^\pm_1\to W^\pm Z)$.
A measurement of the total resonance width $\Gamma(V^\pm_1\to{\rm anything})$
would remove the dependence on the unknown branching fraction $BR$.
However, the accuracy of this measurement is severely limited by the
poor missing energy resolution. Even though a Higgsless origin of the
resonance can be ruled out if the value of $g^{(1)}_{WZV}$,
inferred with the assumption of $BR=1$, violates the bound (\ref{bound}),
the LHC alone will not be able to settle the issue and precise
measurements at the ILC appear to be necessary for the ultimate 
test of the theory. 


\section{COLLIDER PHENOMENOLOGY AT THE ILC}
\label{sec:ilc}

\begin{figure*}[t]
\centering
\includegraphics[width=85mm]{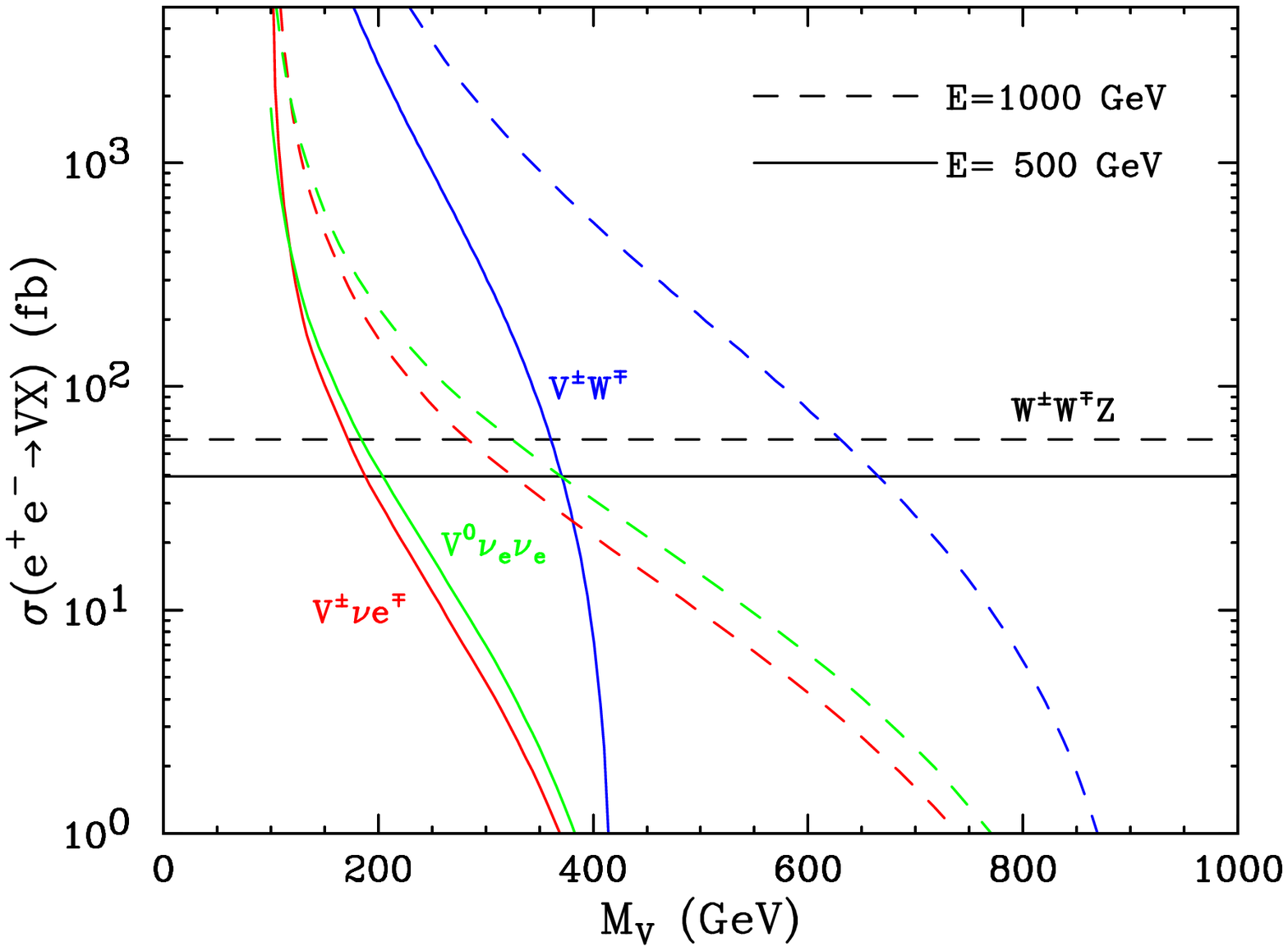}
\includegraphics[width=85mm]{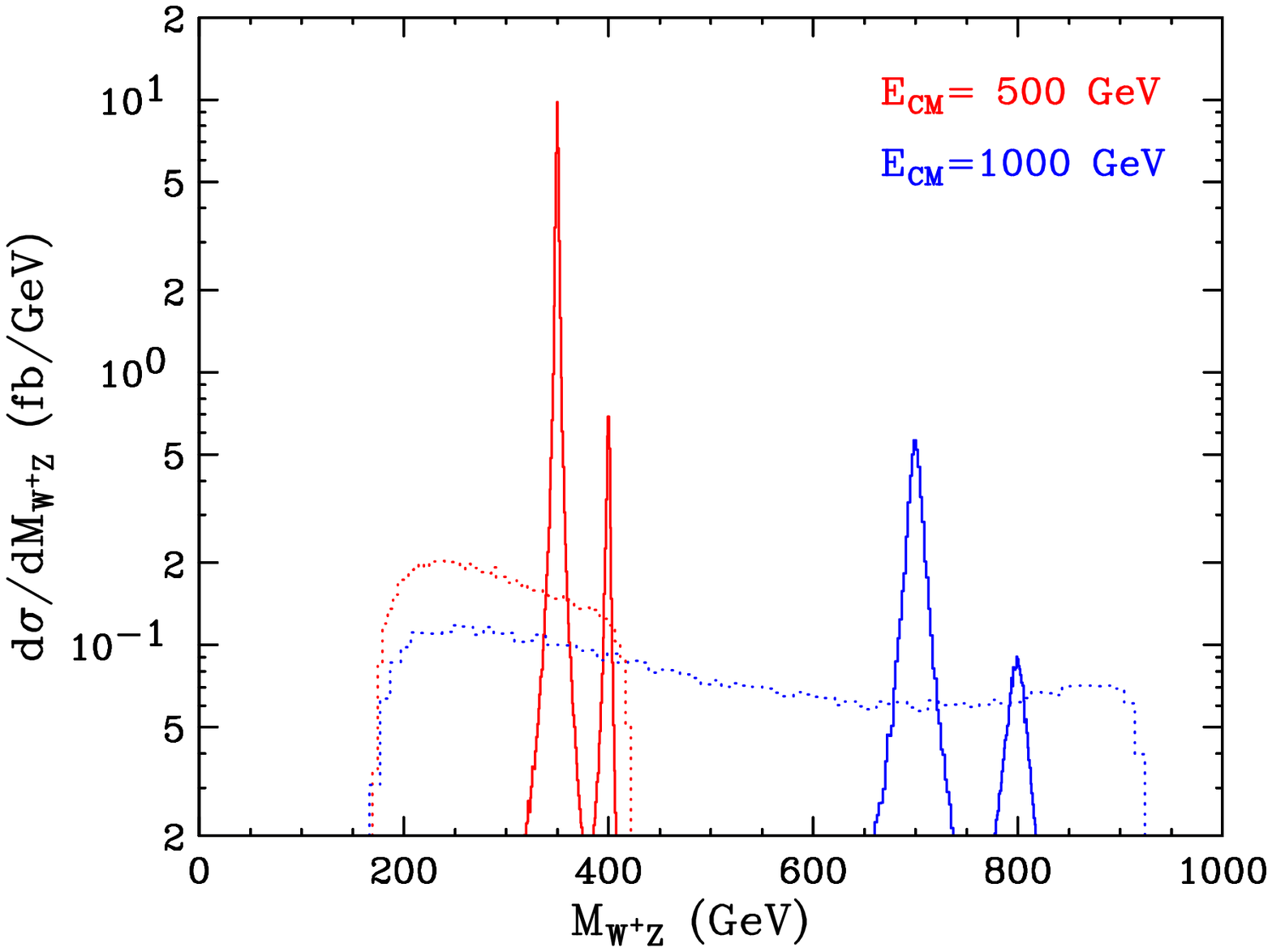}
\caption{Left: $V_1$ production cross-sections and the continuum 
SM background at an $e^+e^-$ 
lepton collider of center of mass energy 500 GeV (solid)
or 1 TeV (dashed). Right: $WZ$ invariant mass distribution for
Higgsless signals (solid) and SM background (dotted), at 
$E_{CM}=500$ GeV (red, $M^\pm=350,400$ GeV) and 
$E_{CM}=1$ TeV (blue, $M^\pm=700,800$ GeV).} \label{fig:ilc}
\end{figure*}

Unlike traditional technicolor, Higgsless models offer new discovery 
opportunities for a lepton collider with a center-of-mass energy in the
sub-TeV range. From \leqn{NDA} we have seen that the masses of the new
MVBs are expected to be below 1 TeV, and they can be produced 
at the ILC through the analogous vector boson fusion process
by bremsstrahlung of $W$'s and $Z$'s off the initial state 
$e^+$ and $e^-$. The $V_1$ production cross-sections for vector boson fusion
$e^+e^-\to V^\pm_1 e^\mp \nu_e$ and $e^+e^-\to V^0_1 \nu_e \bar{\nu}_e$
as well as associated production $e^+e^-\to V^\pm W^\mp$ are shown in 
the left panel of Fig.~\ref{fig:ilc}. The horizontal lines correspond 
to the total cross-section of the continuum SM background. We see that
for a large range of $V_1$ masses, ILC searches appear promising, 
already at the level of total number of events, before cuts and 
efficiencies. Furthermore, because of the cleaner environment of the ILC,
one can now use the dominant hadronic decay modes of the $W$ and $Z$,
and easily reconstruct the invariant mass of the $V_1$ resonance,
which provides an extra handle for background suppression 
(see the right panel in Fig.~\ref{fig:ilc}).
Further detailed studies are needed to better evaluate the 
ILC potential for testing the generic predictions (\ref{sumW}) and (\ref{sumZ})
of the Higgsless models.

\begin{acknowledgments}
MP is supported by NSF grant PHY-0355005. KM and AB are supported by a US DoE
Outstanding Junior Investigator award under grant DE-FG02-97ER41029.
\end{acknowledgments}


\end{document}